\documentclass[prb,preprint]{revtex4-1}
\usepackage{amsmath}
\usepackage{amsfonts}
\usepackage{graphicx}
\usepackage{epstopdf}

\newcommand{\R}{\mathbb R}
\newcommand{\dif}[1]{\textnormal{d}#1}

\begin{document}

\title{A note on a straight gravity tunnel through a rotating body}
\author{Aleksander Simoni\v{c}}
\email{aleks.simonic@gmail.com}
\affiliation{School of Science, The University of New South Wales (Canberra), ACT, Australia}

\begin{abstract}
It is well-known that the straight gravity tunnel between any two different positions on a non-rotating Earth, which has uniform density, is traversable, i.e., an object initially at rest will reach its destination through the gravity tunnel in both directions. Moreover, the time taken to fall is always constant. These facts are no longer true if rotation is allowed. The aim of this note is to derive the necessary and sufficient condition for traversability of straight gravity tunnels through a rotating physical body with spherically symmetric gravitational field. Fall-through times are expressed in a closed form for linear and constant gravitational fields. In conclusion, these models are compared to numerically obtained data using the internal structure of the Earth.
\end{abstract}

\maketitle

\section{Introduction}

The concept of the gravity tunnel refers to the travel along frictionless subterranean passages from place $A$ to place $B$ which are positioned on a surface of a massive physical body, e.g., the Earth, with no expenditure of energy for locomotion. Although the corresponding vehicle called the gravity train is probably in the domain of science fiction only due to obvious engineering problems, this thought experiment is still an active field of research, which also paves the way for undergraduate physics. A classical result by Cooper \cite{Cooper} asserts that a gravity train has a simple harmonic motion and always needs a constant amount of time (roughly $42$ minutes) to fall through a chord path (a straight gravity tunnel) between any two different positions on a non-rotating Earth, which has uniform density, i.e., linear gravitational field. Recently, several authors have discussed other planet's interiors and gravitational fields \cite{Isermann,Klotz,Pesnell}, e.g., constant gravity, the gravitational strength predicted by the Preliminary Reference Earth Model\cite{Dziewonski} (PREM) and polytropes, as well as gravity tunnels with relativistic \cite{Parker} and friction \cite{tunneldrag} effects. The brachistochrone path, i.e., a path which minimizes the time between any two points, is also studied in some of these papers. Some authors refer as gravity tunnels only object's free fall trajectories since then the gravity train is no longer subject to frictional and inertial forces. Results in this direction \cite{remark} are known for rotating homogeneous spheres \cite{SimosonFreeFall}, rotating homogeneous flattened spheroids \cite{Taillet} and rotating PREM Earth \cite{Isermann2}. However, in this note only straight gravity tunnels are considered. Some references to the history of gravity tunnels can be found in Selmke's note \cite{Selmke}.

The most common example of the gravity tunnel is through the center of a non-rotating Earth. Taking angular frequency $\omega_{\oplus}=\sqrt{g_{\oplus}/R_{\oplus}}$ where $g_{\oplus}=9.807~\textrm{m}\textrm{s}^{-2}$ is the gravitational field strength at the surface of the Earth with the radius $R_{\oplus}=6.371\times10^6~\mathrm{m}$, it follows by simple kinematics that the half-period of oscillation in the linear gravitational field is $\pi\omega_{\oplus}^{-1}\approx42.19~\textrm{min}$ while in the constant gravitational field it is $2\sqrt{2}\omega_{\oplus}^{-1}\approx38~\textrm{min}$. Note that Cooper's result about the constant fall-through time $T$ is not true in the constant gravitational field. Moreover, $T$ only depends on the distance of the gravity tunnel from the center of the Earth and $2\sqrt{2}\omega_{\oplus}^{-1}\leq T<\pi\omega_{\oplus}^{-1}$.

Surprisingly, none of the above authors considered the rotating case for straight gravity tunnels. In this note, we would like to emphasize that rotation and the spherically symmetric gravitational field imply traversability of straight gravity tunnels. A traversable gravity tunnel means that the gravity train which initially rests at $A$ will reach its destination $B$ and vice versa. It turns out that there are two necessary and sufficient conditions: the first one asserts that the gravity energy is large enough while the second one asserts that the absolute value of $A$'s latitude is the same as the absolute value of $B$'s latitude. The latter is more important because the first condition is, under reasonable circumstances, always satisfied by constant and linear gravitational fields as well as by the PREM. The gravity tunnels which go through the center of a massive body or are perpendicular to the rotational axis are rather specific.

Although Simoson \cite{SimosonSliding} addresses this problem by providing several examples in linear gravitational fields, he does not state this general condition. In the next section we derive Simoson's equation of motion, and in Sec.~\ref{sec:main.result} the general condition. In Sec.~\ref{sec:times} exact formulas for fall-through times in linear and constant gravitational fields are provided and these results are then compared to the PREM.

\section{The equation of motion}

\begin{figure}[h]
\centering
\includegraphics[scale=0.25]{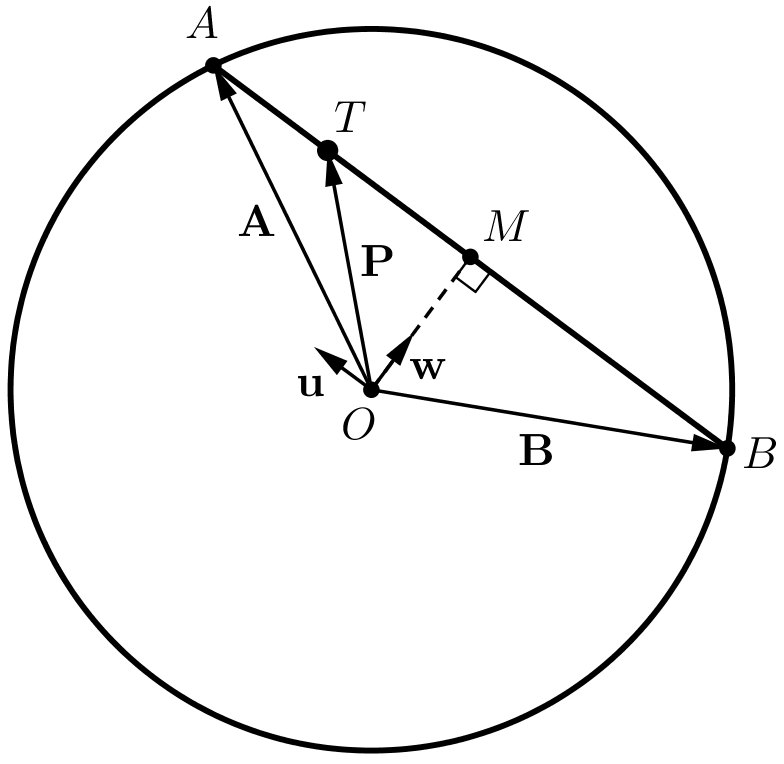}
\caption{The gravity train $T$, which moves along the straight tunnel from $A$ to $B$.}
\label{fig:setting}
\end{figure}

We follow Simoson's paper\cite{SimosonSliding} although our derivation of the equation of motion differs from Simoson's in the sense that we use a non-inertial rotating reference frame instead of applying rotational transformation to inertial one. The reader is advised to consult Fig.~\ref{fig:setting}.

Assume that our physical body is a rotating ball with radius $R$ and rotational period $Q$. Take an orthogonal coordinate system $(x,y,z)\in\R^3$ with the origin $O$ at the center of the ball such that the $z$-axis is the rotational axis. This gives a rise to body's inertial frame $\mathcal{I}=\{O;\mathbf{i},\mathbf{j},\mathbf{k}\}$. Let $x=R\cos{\varphi}\cos{\lambda}$, $y=R\cos{\varphi}\sin{\lambda}$ and $z=R\sin{\varphi}$ be a parametrisation of the surface of this ball with latitude $\varphi\in[-\pi/2,\pi/2]$ and longitude $\lambda\in[-\pi,\pi)$.

Let $A$ and $B$ be two distinct points on the surface of this ball and let $\mathbf{A}$ and $\mathbf{B}$ be corresponding position vectors. Identify the point $A$ with spherical coordinates $\left(\varphi_0,\lambda_0\right)$ and $B$ with $(\varphi,\lambda)$. Take the vector $\mathbf{u}=\left(\mathbf{A}-\mathbf{B}\right)/|\mathbf{A}-\mathbf{B}|$ and let $c=\frac{1}{2}|\mathbf{A}+\mathbf{B}|$ be the distance between $O$ and $M$, the midpoint of the segment $AB$. We have
\begin{gather*}
c = \frac{R}{\sqrt{2}}\sqrt{1+\sin{\varphi}\sin{\varphi_0}+\cos{\varphi}\cos{\varphi_0}\cos{\left(\lambda-\lambda_0\right)}}, \\
\mathbf{u} = \frac{-1}{2\sqrt{1-(c/R)^2}}\left(\cos{\varphi}\cos{\lambda}-\cos{\varphi_0}\cos{\lambda_0},\cos{\varphi}\sin{\lambda}-\cos{\varphi_0}\sin{\lambda_0},
\sin{\varphi}-\sin{\varphi_0}\right).
\end{gather*}
Points $A$ and $B$ are antipodal points if and only if $c=0$. If $c\neq0$, then we take the vector $\mathbf{w}=\left(\mathbf{A}+\mathbf{B}\right)/|\mathbf{A}+\mathbf{B}|$ which is
\[
   \mathbf{w} = \frac{R}{2c}\left(\cos{\varphi}\cos{\lambda}+\cos{\varphi_0}\cos{\lambda_0},\cos{\varphi}\sin{\lambda}+\cos{\varphi_0}\sin{\lambda_0},
\sin{\varphi}+\sin{\varphi_0}\right).
\]
If $c=0$, then we take $\mathbf{w} = -\left(\sin{\varphi_0}\cos{\lambda_0},\sin{\varphi_0}\sin{\lambda_0},-\cos{\varphi_0}\right)$ if $\varphi_0\notin\{-\pi/2,\pi/2\}$, and $\mathbf{w}=(1,0,0)$ otherwise. Observe that vectors $\mathbf{u}$ and $\mathbf{w}$ are always the unit vectors and $\mathbf{u}$ is parallel to the line $AB$ while $\mathbf{w}$ is perpendicular to it. We can take $\mathcal{F}=\left\{O;\mathbf{u},\mathbf{w},\mathbf{u}\times\mathbf{w}\right\}$ as a non-inertial frame which rotates with the angular velocity $\mathbf{\Omega}=(2\pi/Q)\mathbf{k}$ relative to $\mathcal{I}$. Then one can describe the position of the gravity train $T$ in the frame $\mathcal{F}$, which moves along the straight tunnel from $A$ to $B$ as a function $\mathbf{P}(t)=c\mathbf{w}+s(t)\mathbf{u}$ dependent on time $t$ such that $\mathbf{P}(0)=\mathbf{A}$, i.e., $s(0)=\frac{1}{2}|\mathbf{A}-\mathbf{B}|$. By the acceleration transformation formula\cite{GregoryCM} we have
\begin{equation}
\label{eq:generaleq}
\ddot{\mathbf{P}} = -\frac{f\left(\left|\mathbf{P}\right|\right)}{\left|\mathbf{P}\right|}\mathbf{P} + \left(\mathbf{\Omega}\times\mathbf{P}\right)\times\mathbf{\Omega} + 2\dot{\mathbf{P}}\times\mathbf{\Omega} + \mathbf{P}\times\dot{\mathbf{\Omega}}
\end{equation}
where $f(r)$, $0\leq r\leq R$ is a spherically symmetric gravitational field of the physical body. The Euler acceleration $\mathbf{P}\times\dot{\mathbf{\Omega}}$ is obviously zero. After multiplying both sides of Eq.~\eqref{eq:generaleq} by $\mathbf{u}$, it is easy to see that $2\left(\dot{\mathbf{P}}\times\mathbf{\Omega}\right)\mathbf{u}=0$ and
\[
   \left(\left(\mathbf{\Omega}\times\mathbf{P}\right)\times\mathbf{\Omega}\right)\mathbf{u} = \left|\mathbf{\Omega}\right|^2\mathbf{P}\mathbf{u}-\left(\mathbf{\Omega}\mathbf{P}\right)\left(\mathbf{\Omega}\mathbf{u}\right) =
   \left(\frac{2\pi}{Q}\right)^2\left(s(t)\left(1-u_3^2\right)-cu_3w_3\right).
\]
Therefore, the only fictitious acceleration which plays a role here is the centrifugal acceleration. Thus the function $s(t)$ must satisfy the equation
\begin{equation}
\label{eq:simoson}
\ddot{s} = \left(-\frac{f\left(\sqrt{c^2+s^2}\right)}{\sqrt{c^2+s^2}}+\left(\frac{2\pi}{Q}\right)^2\left(1-u_3^2\right)\right)s - c\left(\frac{2\pi}{Q}\right)^2u_3w_3.
\end{equation}
In the next section we will write Eq.~\eqref{eq:simoson} in a dimensionless form.

\section{The main result}
\label{sec:main.result}

In order to write Eq.~\eqref{eq:simoson} in a dimensionless form with coefficients expressed in spherical coordinates, take $s(t)=R\chi(\tau)$, $t=Q(2\pi)^{-1}\tau$, $F(\chi)=Q^2R^{-1}(2\pi)^{-2}f\left(R\chi\right)$ and $\bar{c}=cR^{-1}$. Observe that $F(\chi)$ is a positive function defined on the interval $[0,1]$. Using expressions for $\mathbf{u}$ and $\mathbf{w}$ from the previous section, Eq.~\eqref{eq:simoson} becomes
\begin{equation}
\label{eq:simoson2}
\frac{\dif{}^2\chi}{\dif{\tau}^2} = \left(-\frac{F\left(\sqrt{\chi^2+\bar{c}^2}\right)}{\sqrt{\chi^2+\bar{c}^2}}+a\right)\chi + b
\end{equation}
with real numbers
\begin{equation*}
a = 1-\frac{\left(\sin{\varphi}-\sin{\varphi_0}\right)^2}{4\left(1-\bar{c}^2\right)}, \quad
b = \frac{\sin{\left(\varphi+\varphi_0\right)}\sin{\left(\varphi-\varphi_0\right)}}{4\sqrt{1-\bar{c}^2}}.
\end{equation*}
Formula for $b$ is correct also in the case $c=0$ since then $\varphi=-\varphi_0$ and $b$ is zero, as it must be due to the last term in Eq.~\eqref{eq:simoson}.

According to the definitions, we have $0\leq\bar{c}<1$ and $0\leq a\leq1$. Define $\bar{g}=\omega^2Q^2(2\pi)^{-2}$ where $\omega=\sqrt{g/R}$ and $g=f(R)$ is the gravitational field strength at the surface of the body. For the Earth there is $\omega_{\oplus}=1.241\times10^{-3}~\textrm{s}^{-1}$ and $Q_{\oplus}=8.6164\times10^{4}~\textrm{s}$, therefore $\bar{g}_{\oplus}=289.5$. As mentioned before, the popular choices for $F$ are the constant gravitational field $F_{\mathrm{con}}=\bar{g}$ and the linear gravitational field $F_{\mathrm{lin}}(\chi)=\bar{g}\chi$. Here $F_{\oplus}(\chi)$ denotes the gravitational field predicted by the PREM.

Assume that the gravity train is at rest at $A$. It is possible to integrate Eq.~\eqref{eq:simoson2} to obtain
\begin{equation}
\label{eq:time}
\left(\frac{\dif{\chi}}{\dif{\tau}}\right)^2 = \int_{\sqrt{\chi^2+\bar{c}^2}}^{1} 2F(u)\dif{u} - a\left(1-\chi^2-\bar{c}^2\right) + 2b\left(\chi-\sqrt{1-\bar{c}^2}\right).
\end{equation}
The gravity tunnel is traversable if and only if the right-hand side of Eq.~\eqref{eq:time} is, as a function of $\chi$, positive for $-\sqrt{1-\bar{c}^2}<\chi<\sqrt{1-\bar{c}^2}$ and equals zero for $\chi=-\sqrt{1-\bar{c}^2}$. While the first condition ensures that the gravity train reaches $B$, the second condition means that it also stops there. The latter condition which is equivalent to $b=0$ or $|\varphi|=|\varphi_0|$ is crucial for the gravity train to be able to move back to $A$. We thus have two necessary and sufficient conditions for traversability:
\begin{enumerate}
\item For $0\leq \chi<\sqrt{1-\bar{c}^2}$ it follows $$\int_{\sqrt{\chi^2+\bar{c}^2}}^{1}F(u)\dif{u} > \frac{a}{2}\left(1-\chi^2-\bar{c}^2\right);$$
\item $|\varphi|=|\varphi_0|$.
\end{enumerate}
The first condition is always satisfied in the case of constant and linear gravitational fields until $\bar{g}>1$. This is taken for granted in what follows. The PREM model also satisfies this condition since $F_{\oplus}(\chi)$ is always greater than $F_{\mathrm{lin}}(\chi)$. Therefore, the second condition is crucial. For the sake of simplicity, we say that the gravity tunnel is \emph{horizontal} if $\varphi=\varphi_0$ and \emph{vertical} if $\varphi=-\varphi_0$. Throughout this note, these two types of gravity tunnels are of interest to us since we are only interested in traversable gravity tunnels.

From Eq.~\eqref{eq:time} the general expression for the fall-through time is deduced:
\begin{equation}
\label{eq:arr.time}
T = \frac{2\sqrt{\bar{g}}}{\omega}\int_{\bar{c}}^1 w\left(\left(w^2-\bar{c}^2\right)\left(\int_{w}^{1}
2F(u)\dif{u}+a\left(w^2-1\right)\right)\right)^{-\frac{1}{2}}\dif{w}.
\end{equation}
This integral depends on $\varphi_0$ and $\bar{c}$ because $\varphi=\pm\varphi_0$ and $a$ depends only on $\varphi_0$ and $\bar{c}$. If by $\theta_0$ the angle between vectors $\mathbf{B}$ (or $\mathbf{A}$) and $\mathbf{w}$ is denoted, then $\bar{c}=\cos{\theta_0}$. Therefore, $\bar{c}$ has a similar role as $\theta_0$ in Klotz's paper. For the horizontal gravity tunnel we have $a=1$ and $T$ really depends only on $\bar{c}\in\left[\sin{\left|\varphi_0\right|},1\right]$. But for the vertical gravity tunnel we have $a=1-\left(1-\bar{c}^2\right)^{-1}\sin^2{\varphi_0}$ and $\varphi_0$ must be taken into account where $\bar{c}\in\left[0,\cos{\varphi_0}\right]$. Of course, $T$ is an even function in variable $\varphi_0$.

\section{Fall-through times}
\label{sec:times}

In this section, exact formulas for fall-through times in linear and constant gravitational fields are provided. In general, these equations depend on variables $a$ and $\bar{c}$ and on parameters $\bar{g}$ and $\omega$. Observe that the limiting case $\bar{g}\to\infty$ corresponds to the non-rotating scenario. This means that we must obtain all non-rotating formulas after taking this limit in equations below. At the end, we compare both models with the PREM.

\subsection{Linear gravitational field}

If we take $F(\chi)=F_{\mathrm{lin}}(\chi)$, then Eq.~\eqref{eq:arr.time} gives
\[
   T = \frac{\pi}{\omega}\sqrt{\frac{1}{1-a/\bar{g}}}.
\]
In this case the fall-through time is constant for horizontal gravity tunnels only but changes a little if $\bar{g}$ is sufficiently large. For the Earth, this value is $42.26~\textrm{min}$, slightly more than for a non-rotating Earth.

\subsection{Constant gravitational field}

If we take $F=F_{\mathrm{con}}$, then it is possible to find an equation for $T$ in terms of complete elliptic integrals. From Eq.~\eqref{eq:arr.time} it follows
\[
   T = \frac{\sqrt{\bar{g}}}{\omega}\int_{\bar{c}}^1 \frac{2w \dif{w}}{\sqrt{(a(w+1)-2\bar{g})(w-1)(w-\bar{c})(w+\bar{c})}}.
\]
In order to write this integral in a closed form, we need to consider three separate cases: $a>0$ and $\bar{c}>0$, $a>0$ and $\bar{c}=0$, and $a=0$. Accordingly, we have:
\begin{enumerate}
\item
\begin{flalign}
\label{eq:time1}
T = \frac{4\bar{c}}{\omega\sqrt{(1+\bar{c})(2-a(1+\bar{c})/\bar{g})}}\Biggl(&2\mathbf{\Pi}\left(\frac{1-\bar{c}}{1+\bar{c}},
   \sqrt{\frac{(1-\bar{c})(2-a(1-\bar{c})/\bar{g})}{(1+\bar{c})(2-a(1+\bar{c})/\bar{g})}}\right) \nonumber \\
   &-\mathbf{K}\left(\sqrt{\frac{(1-\bar{c})(2-a(1-\bar{c})/\bar{g})}{(1+\bar{c})(2-a(1+\bar{c})/\bar{g})}}\right)\Biggr);
\end{flalign}
\item
\begin{equation}
\label{eq:time3}
T = \frac{2}{\omega}\sqrt{\frac{\bar{g}}{a}}\log{\frac{\bar{g}-a}{\bar{g}-\sqrt{a\left(2\bar{g}-a\right)}}};
\end{equation}
\item
\begin{equation}
\label{eq:time2}
T = \frac{2\sqrt{2}}{\omega\sqrt{1+\bar{c}}}\left(\left(1+\bar{c}\right)\mathbf{E}\left(\sqrt{\frac{1-\bar{c}}{1+\bar{c}}}\right)
-\bar{c}\mathbf{K}\left(\sqrt{\frac{1-\bar{c}}{1+\bar{c}}}\right)\right),
\end{equation}
\end{enumerate}
respectively. Here $\mathbf{K}$, $\mathbf{E}$ and $\mathbf{\Pi}$ are the first, second and third complete elliptic integrals \cite{TableIntegrals}.
Eqs.~\eqref{eq:time3} and \eqref{eq:time2} can be obtained from Eq.~\eqref{eq:time1} after taking limits $\bar{c}\to0$ and $a\to0$, respectively.
Eq.~\eqref{eq:time2} is the limiting case of Eq.~\eqref{eq:time1} after $\bar{g}\to\infty$ and thus also presents the non-rotating fall-through time. This equation is the same as Klotz's Eq.~A10 except that in his equation a minus must be used between elliptic integrals, the same remark made also by Isermann \cite{Isermann} who obtained analytic expressions for fall-through times in the case of a non-rotating Earth with piecewise linear approximation of the PREM.

\subsection{The PREM model}

\begin{figure}[h]
\centering
\includegraphics[scale=0.1]{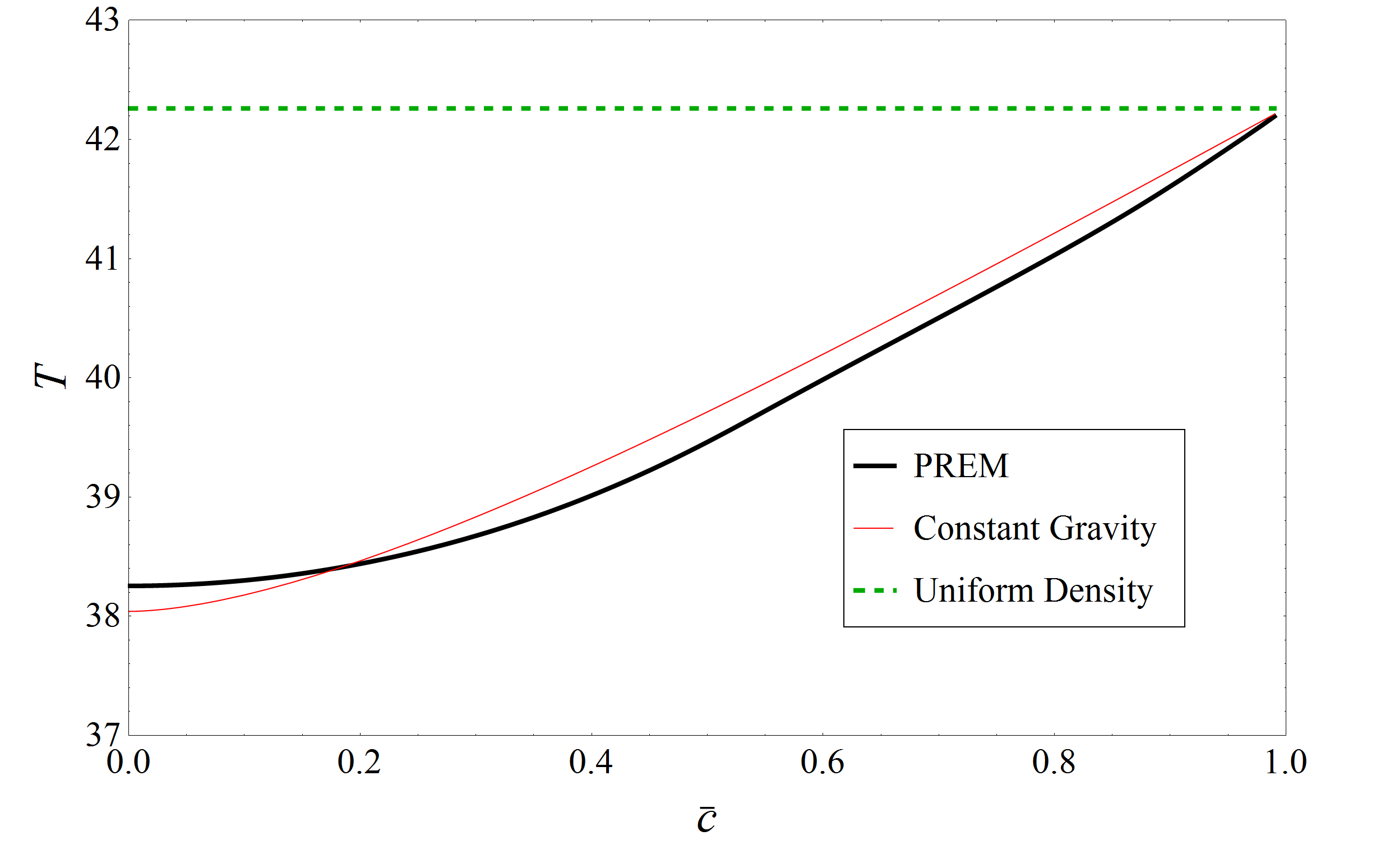}
\caption{Fall-through time $T$ (in minutes) as a function of distance of the horizontal gravity tunnel from the Earth's center per the Earth's radius $\bar{c}$ according to the PREM, constant gravity and uniform density models.}
\label{fig:GravTunnel1}
\end{figure}

\begin{figure}[h]
\centering
\includegraphics[scale=0.15]{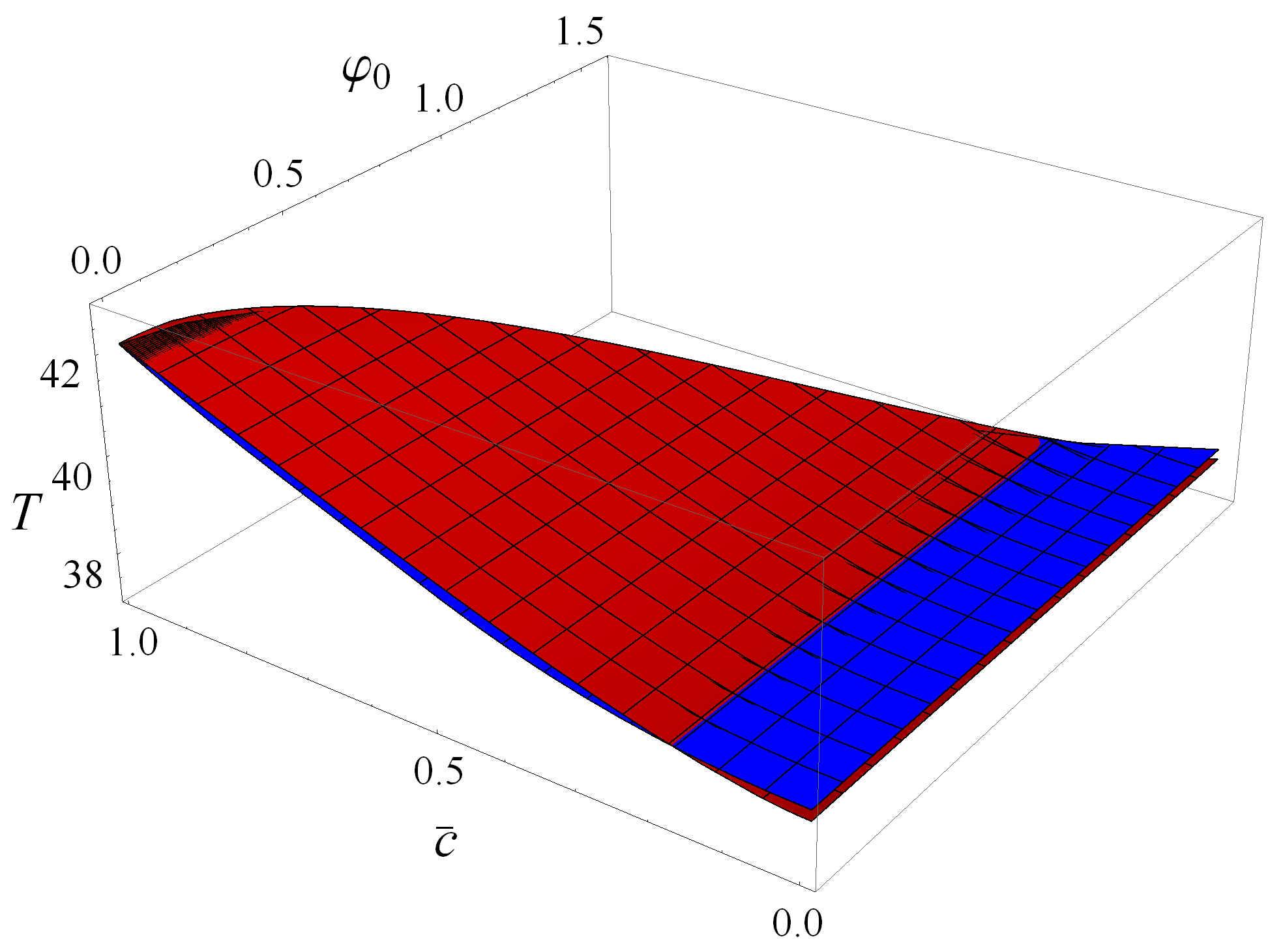}
\caption{Fall-through time $T$ (in minutes) as a function of distance of the vertical gravity tunnel from the Earth's center per the Earth's radius $\bar{c}$ and departure latitude $\varphi_0$ (in radians) according to the PREM (the blue surface) and constant gravity (the red surface) models.}
\label{fig:GravTunnel2}
\end{figure}

Klotz used the PREM to get fall-through times depending on $\theta_0$. Interestingly, the PREM results are much closer to the constant gravitational field model than the linear gravitational field model although the latter is more physically realistic concerning gravity around the origin. Our results for a rotating Earth are, unsurprisingly, essentially the same, see Fig.~\ref{fig:GravTunnel1} and Fig.~\ref{fig:GravTunnel2}.

\section{Conclusion}

The point of this note is that we must have traversable gravity tunnels in order to be able to speak about fall-through times. While this is irrelevant for non-rotating physical bodies, it is crucial in other cases. In this note, we have investigated the necessary conditions for straight gravity tunnels to be traversable. One of them greatly narrows the choice of allowed positions of tunnel's endpoints on a massive body; the absolute values of endpoints' latitudes must be the same. We compare the theoretically obtained fall-through times in linear and constant gravitational fields for the Earth with numerics of the PREM. The results are close to the non-rotating times because the Earth rotates relatively slowly. If we want the rotation to have a great impact, then $\bar{g}$ must be close to $1$. This means that the Earth should rotate close to the period $1.4~\mathrm{h}$.

\begin{acknowledgments}
We would like to thank Nada Razpet and Mitja Rosina for their helpful remarks while writing the manuscript.
\end{acknowledgments}



\begin{thebibliography}{99}
\bibitem{remark} Interesting thought experiment due to Hawking is dropping a small black hole through the rotating Earth. One of natural questions is what happens when the black hole starts gobbling the Earth at a set constant rate, see Simoson's note \cite{SimosonBlackHole}.
\end{thebibliography}
\end{document}